\documentclass[aps,twocolumn,prl,prlshowpacs,groupedaddress,nofootinbib]{revtex4}  % for review and submission 
\usepackage{graphicx}  % needed for figures
\usepackage{dcolumn}   % needed for some tables
\usepackage{bm}        % for math
\usepackage{amssymb}   % for math
\usepackage{slashed}
\usepackage{color}
\usepackage{lipsum}
\usepackage{textcomp}
\usepackage{appendix}
\usepackage{CJKutf8}
\usepackage{slashed}
\usepackage{lipsum} % 调用跨栏长公式宏包
\begin{document}	
%\begin{CJK}{UTF8}{gbsn}	
	\title{A unified scheme for calculating the exclusive semi-leptonic decays of hadrons}	
	%一种计算强子发生遍举半轻衰变的统一方案
	\author{Guo-He Yang} \email{ygh@hnit.edu.cn}
\affiliation{Hunan Institute of Technology, HENGYANG, China}
%College of science, Hunan Institute of  Technology, Hunan 421002, China 

	\date{\today}
	\providecommand{\keywords}[1]{\textbf{\textit{keywords---}} #1}	
	\begin{abstract} 	
	Exclusive semi-leptonic decays are a pivotal channel for exploring heavy-flavor physics, primarily because they are straightforwardly measurable experimentally. In this work, we delve into the hadron matrix element and hadron tensor within the context of exclusive semi-leptonic decays. We challenge the conventional exclusive decay theory by introducing a fresh perspective, revealing that while the baryon sector is consistent, the meson sector warrants revision. Employing a novel form factor derived from the Taylor series expansion in our differential-width calculations, we demonstrate that fitting experimental data requires a more streamlined, minimal-parameter set than the standard Light Cone Sum Rule (LCSR) form factors. Furthermore, we propose that our hypothesis can be empirically validated or refuted by precise measurement of the double-differential decay width, providing a tangible path forward for experimental validation.
	% We have also generalized the new method to the conditions where the strict V-A process is not satisfied and $( f_2, g_2, f_3, g_3) \not=0$.
	%遍举半轻衰变因其实验测量简单而成为重味物理研究的重要途径。本文研究了遍举半轻衰变中的强子矩阵元和强子张量。我们从一个新的角度重新思考现有的遍举衰变理论，结果是重子部分没有问题，但介子部分需要修正。采用泰勒级数展开新形状因子计算的微分宽度拟合实验数据，所需参数比普通LCSR形状因子不仅少而且更简单。我们也提出了实验上可以通过测量双微分衰变宽度来验证或推翻我的假设。
\end{abstract}
%semi-leptonic decay; inclusive decay; exclusive decay; form factor;decay width
\maketitle

\section{Introduction}
The heavy quark expansion (HQE) in powers of $1/m_Q  $ has long been the cornerstone of the standard theoretical framework for analyzing the decays of bottom and charmed hadrons. However, this approach is currently grappling with significant challenges, particularly in explaining the lifetimes of charmed baryons. In July 2018, the LHCb Collaboration reported  a measurement \cite{LHCb:2018nfa}of the $ \Omega_c^0 $ baryon's lifetime, which is nearly four times longer than the 2018 world average value\cite{ParticleDataGroup:2018ovx}.This finding was reaffirmed by the LHCb experimental group in 2021 \cite{LHCb:2021vll}. This experimental finding is difficult to understand under the current theoretical framework, and even a qualitative explanation cannot be provided. Another formidable challenge lies in the precise measurement of the CKM parameters
$ V_{cb} $ and $ V_{ub} $. The inclusive measurements suggest averages of 
$|V_{cb}|=(42.2\pm0.8)\times10^{-3}$, and $|V_{ub}|=(4.13\pm0.12^{+0.13}_{-0.14}\pm0.18)\times10^{-3}$, while exclusive measurements yield  $|V_{cb}|=(39.4 \pm 0.8)\times10^{-3},|V_{ub}|=(3.70\pm0.10\pm0.12)\times10^{-3}$\cite{ParticleDataGroup:2022pth}.	As experimental precision improves, the divergence between these two methodologies has become increasingly pronounced.  The experimental average of $R(D)$ and $R(D^{*})$, which deviate from Standard Model predictions \cite{HFLAV:2022esi}, underscore a critical issue in heavy flavor physics. These inconsistencies suggest a flaw within the theoretical models. I contend that the indiscriminate addition of higher-order operators and high-loop diagram corrections is not a panacea for these contradictions. Instead, a reevaluation of our existing theoretical frameworks is imperative.
%$1/m_Q  $次方的重夸克展开(HQE)是分析底强子和粲强子衰变的标准理论框架。目前，它对粲重子的寿命问题面临着一个巨大的挑战。在PDG2018版本\cite{ParticleDataGroup:2018ovx}中，从固定靶实验中提取的世界平均值为$ \tau(\Omega_c^0)= (0.69 ± 0.12) × 10^{-13} $， 和粲重子寿命顺序是$ \tau(\Xi_c^+)>\tau(\Lambda_c^+)>\tau(\Xi_c^0)>\tau(\Omega_c^0)$。LHCb合作组报告了2018年7月使用衰变通道$ \Omega_b^-\rightarrow\Omega_c^0\mu^-\bar{\nu}_\mu X$和$\Omega_c^0\rightarrow pK^-K^-\pi^+  $对$ \Omega_c^0 $重子$  \tau(\Omega_c^0)=(2.68\pm0.24\pm0.10\pm0.02)\times10^{-13}s$寿命的测量\cite{LHCb:2018nfa}。它几乎是2018年世界平均价值的四倍。2021年，LHCb实验组在$\Omega_c^0$上用最新的测量结果证实了之前基于半轻底强子衰变的LHCb测量结果，并将实验精度提高了两倍\cite{LHCb:2021vll}。这一实验发现在目前的理论框架下很难理解，甚至无法进行定性解释。	
%另一个挑战是CKM参数的测量$ V_{ub} $和$ V_{cb} $。单举测量值的平均值由$|V_{cb}|=(42.2\pm0.8)\times10^{-3},|V_{ub}|=(4.13\pm0.12^{+0.13}_{-0.14}\pm0.18)\times10^{-3}$提供。遍举测量的平均值由$|V_{cb}|=(39.4 \pm 0.8)\times10^{-3},|V_{ub}|=(3.70\pm0.10\pm0.12)\times10^{-3}$\cite{ParticleDataGroup:2022pth}提供。随着实验精度的提高，两者之间的偏差越来越明显。实验结果的不一致一定意味着这个理论有问题。我认为，不加区分地添加高阶运算符和高环路图修正并不是解决这些矛盾的灵丹妙药。相反，重新评估我们现有的理论框架是必要的。	

The inspiration for this article emerged from my reflections on inclusive decay during my learning process, and I intend to write an article on the theory of inclusive decay in the future. For now, however, let us concentrate on the more tractable issue of exclusive decay. The fundamental divergence between my approach and traditional methods lies in the treatment of hadron matrix elements.

%这篇文章的灵感是我在学习和思考单举衰变的过程中产生的，以后我也会写一篇关于单举衰变理论的文章。现在，让我们关注更易于掌握的遍举衰变。我的方法与传统方法最大的不同在于如何处理强子矩阵素。

The rest of the paper is organized as follows. In Section 2, we introduce a unified scheme for calculating the exclusive decay width of hadron semi-leptonic decay.  We calculate the differential decay width and moment by a new method in Section 3. In Section 4, we compare our method with the traditional theory of baryon decay, and it turns out that the result is totally compatible. In Section 5, we compare our method with the traditional theory of meson decay, and we demonstrate that a vector made of spinors is not necessarily a combination of four momentum. In Section 6, we apply the new method to the decay of pseudoscalar mesons into pseudoscalar mesons. By experimentally measuring the distribution of the double differential width, it can be determined which theory is the best. We conclude with Section 7.
%本文的其余部分组织如下。在第二节中，我们提出了一种计算强子半轻衰变的遍举衰变宽度的新方法。在第3节中，我们用一种新的方法计算微分衰减宽度和矩。在第4节中，将我们的方法与传统的重子衰变理论进行比较，结果是完全相容的。在第5节中，将我们的方法与传统的介子衰变理论进行比较，我们表明由旋量组成的矢量不一定是四动量的组合。第6节中我们将新方法应用到赝标量介子到赝标量介子的衰变中，通过实验测量双微分宽度的分布，可以明确到底哪个理论才是最佳的。我们以第7节结束。	
\section{Put forward new ideas}%提出新想法
Consider the decay process $h_i\to h_f+e^-+v_e $ as a case in point, with the corresponding quark-level transition being $b\to c+e^-+v_e $.  By the optical theorem, the imaginary part of the forward scattering amplitude is given by $2 Im \mathcal M(i\to i)=\int d\Pi_f|\mathcal M( i \to f)|^2 $, where $ d\Pi_f=\frac{d^{3}p_{k}}{(2\pi)^{3}2E_{k}}$ represents the phase space of all final state particles.  The imaginary part of the forward scattering amplitude is 
%我们以$h_i\to h_f+e+v_e $为例，夸克层次的衰变是$b\to c+e^-+v_e $。根据光学定理$2 Im \mathcal M(i\to i)=\sum_f \int d \Pi_f|\mathcal M( i \to f)|^2 $,其中$\Pi_f$代表对所有的末态粒子的相空间积分。矩阵元的虚部是
\begin{eqnarray}
	\label{constant}
	&&2Im \mathcal{M}(h_i\to h_i) \nonumber \\
	%&=&\int[d^3p_e][d^3p_{\nu_e}][d^3p_{h_f}]h_{\mu\nu}L^{\mu\nu}(p_e,p_{\nu_e})\nonumber\\&&\left(2\pi\right)^4\delta^4(p_{h_i}-p_{h_f}-p_{\nu_e}-p_e)\nonumber\\
	&=&\int\frac{dQ^2}{2\pi}\int[d^3q][d^3p_{h_f}]h_{\mu\nu}\left(2\pi\right)^4\delta^4(p_{h_i}-q-p_{h_f})\nonumber\\&&\int[d^3p_e][d^3p_{\nu_e}]L^{\mu\nu}(p_e,p_{\nu_e})\left(2\pi\right)^4\delta^4(q-p_e-p_{\nu_e})\;,\quad
\end{eqnarray}	
where the momentum of the initial hadron, final hadron, electron, neutrino, and lepton are $  p_{h_i}$, $p_{h_f}$, $p_e$, $ p_{v_e}$, and $ q$, respectively. $q=p_e+ p_{v_e}=p_{h_i}-p_{h_f}$, the squared of four momenta $Q^2=q^2$. The hadron tensor is
\begin{eqnarray}
	h_{\mu\nu}=\left \langle h_i|\bar{b}\gamma^{\mu}P_{L}c | h_f\right\rangle\left \langle h_f|\bar{c}\gamma^{\nu}P_{L}b| h_i\right\rangle\;, 	
\end{eqnarray} 
and the leptonic tensor is
%其中初始强子、最终强子、电子和中微子的动量分别为$  p_{h_i},p_{h_f}, p_e$和$ p_{v_e}$。强子张量是$h_{\mu\nu}=\left \langle h_i|\bar{b}\gamma^{\mu}P_{L}c | h_f\right\rangle\left \langle h_f|\bar{c}\gamma^{\nu}P_{L}b| h_i\right\rangle  $，轻子张量是
\begin{eqnarray}
	L^{\mu\nu}=2\left(p_e^\mu p_{\nu_e}^\nu+p_e^\nu p_{\nu_e}^\mu-g^{\mu\nu}p_e\cdot p_{\nu_e}+ i\varepsilon^{\mu\nu\alpha\beta}p_{e\alpha}p_{\nu_e\beta}\right).\quad 	
\end{eqnarray}
For simplicity, we have elected to omit an overall constant $ 8G_F^2|V_{bc}|^2 $ in eq.(\ref{constant}), and  we have disregarded the masses of the leptons in our calculations. 
The most general form of hadron tensor $h_{\mu\nu}$ is
\begin{eqnarray}
&	h_{\mu\nu}=&-W_1m_{h_i}^2g_{\mu\nu}+W_2p_{\mu}p_{\nu}-iW_3\epsilon_{ \mu \nu\alpha \beta }p^{\alpha}q^{\beta}\nonumber\\
&	&+W_4q_{\mu}q_{\nu}+W_5(p_{\mu}q_{\nu}+p_{\nu}q_{\mu})\;, 	\label{12345}
\end{eqnarray} where $W_i$ are functions of $\hat{q^2}$, $\hat{q^2 }=q^2/m_{h_i}^2$. I propose a novel approach that calculates the double differential width of three-body semi-leptonic decays based on the hadronic tensor in eq.(\ref{12345}), rather than the traditionally used hadronic transition matrix elements. The rationale behind this is that the hadronic tensor can more directly reflect the results of experimental observations, thereby providing us with an analytical method that is more closely aligned with experimental data.

Using cutting rules and integrating leptons’ phase space, we can express the imaginary part of the matrix element as
%为简单起见，我们在式(1)中省略了一个总常数$ 8G_F^2V_{bc}^2 $，并忽略轻子的质量。利用切割规则和对轻子相空间进行积分，我们可以将总速率表示为
\begin{eqnarray}
	&&2Im \mathcal{M}(h_i\to h_i) \nonumber \\
	&	= &\int \frac{d Q^2}{24\pi^2}Im[i\int  \frac{d^{4} q} {(2 \pi)^{4}} \frac{1}{q^{2}-Q^{2}+i \epsilon} \frac{1}{p_{h_f}^{2}-m_{h_f}^{2}+i \epsilon} \nonumber \\&& h_{\mu\nu}( q^{\mu}q^{\nu}-g^{\mu\nu}q^2)]\;.
\end{eqnarray} 

Before learning something new, let's review what we learned from free quark decay.
%1.The inner product of a state vector is a number, the outer product is a matrix, and the sum of the inner product is equal to the sum of its outer product trace.
%在学习新东西之前，让我们回顾一下我们从自由夸克衰变中学到的东西1. 状态向量的内积是一个数字，外积是一个矩阵，内积的和等于它的外积迹的和。
%$\sum_i \left \langle \psi_i|\psi_i\right\rangle=tr[\sum_i \left | \psi_i \right \rangle \left \langle \psi_i \right |]$\\
\begin{eqnarray}
		\left \langle b(p_b,s)|\bar{b}\gamma^{\nu}b| b(p_b,s)\right\rangle&=&\bar{u}(p_b,s)\gamma^{\nu}u(p_b,s)=2P_{b}^{\nu} ,\nonumber \\ 
		\left \langle b(p_b,s)|\bar{b}b| b(p_b,s)\right\rangle&=&\bar{u}(p_b,s)u(p_b,s)=2m_{b},\nonumber \\
		\sum_{spin}b\left | b(p_b,s)\right\rangle \left \langle b (p_b,s)\right |\bar{b}&\to&\sum_{spin}u(p_b,s)\bar{u}(p_b,s)\nonumber \\ &=& \slashed p_b+m_b.
\end{eqnarray}
According to the idea of analogy, we can get
%根据类比的思想，我们可以得到
\begin{eqnarray}
&	&\left \langle h_i|\bar{b}\gamma^{\nu}b| h_i\right\rangle=\bar{u}_{\psi}\gamma^{\nu}u_{\psi },\left \langle h_i|\bar{b}b| h_i\right\rangle=\bar{u}_{\psi }u_{\psi}\;,\nonumber \\
&	&\sum_{spin}b\left |h_i \right\rangle \left \langle h_i \right |\bar{b}\to\sum_{spin} u_{\psi}\bar{u}_{\psi }\;,
\end{eqnarray} where $u_{\psi}  $ is an unknown column vector with four rows and one column. It is divided into two types, corresponding to the quark's spin up and down. For hadron decay, the following trace is a double trace for spinor space and Hilbert space. There exists and can only exist a single independent scalar function $ q^2 $, as indicated by the equations $ 2q \cdot p_{h_i} = m_{h_i}^2 - m_{h_f}^2 + q^2 $ and $ 2q \cdot p_{h_f} = m_{h_i}^2 - m_{h_f}^2 - q^2 $.
%$u_{\psi} $是一个未知的列向量，分为两种类型，对应于夸克的自旋向上和向下。对于强子衰变，求迹是旋量空间和希尔伯特空间的双求迹。有且只有一个独立标量函数q^2$,因为$ 2q\cdot p_{h_i}=m_{h_i}^2-m_{h_f}^2+q^2,2q\cdot p_{h_f}=m_{h_i}^2-m_{h_f}^2-q^2$。
\begin{eqnarray}
   \sum_{spin}	b\left |h_i \right\rangle \left \langle h_i \right |\bar{b}&\to&(X_{b1}\slashed p_{h_i}+X_{b2}\slashed q+X_{b3}m_{h_i}),\nonumber\\ \left \langle h_i|\bar{b}b| h_i\right\rangle&=&2m_{h_f}X_{b3}\;,
\end{eqnarray}   
   where $X_{b1},X_{b2},X_{b3}  $ is the scalar function of $\hat{q^2 }$. We can do something to reduce the number of scalar functions. On the one hand,  
   %式中$X_{b1},X_{b2},X_{b3}  $为$\hat{q^2 }$, $\hat{q^2 }=\frac{q^2}{m_{h_i}^2}$的标量函数。我们可以做一些事情来减少标量函数的数量。一方面，
\begin{eqnarray}
	&&\sum_{spin}	b\left |h_i \right\rangle \left \langle h_i \right |\bar{b}b\left |h_i \right\rangle \left \langle h_i \right |\bar{b}\nonumber \\
	%&=&(X_{b1}\slashed p_{h_i}+X_{b2}\slashed q+X_{b3}m_{h_i})(X_{b1}\slashed p_{h_i}+X_{b2}\slashed q+X_{b3}m_{h_i})\nonumber 	\\
	&\to&[(X_{b1}^2+X_{b3}^2-X_{b1}X_{b2})m_{h_i}^2+(X_{b2}^2-X_{b1}X_{b2})q^2+\nonumber \\&&X_{b1}X_{b2}m_{h_i}^2+2X_{b3} m_{h_i}(X_{b1}\slashed p_{h_i}+X_{b2}\slashed q)]\;.
\end{eqnarray}
On the other hand,
\begin{eqnarray}
	&&\sum_{spin}(b\left |h_i \right\rangle \left \langle h_i \right |\bar{b})*\left \langle h_i|\bar{b}b| h_i\right\rangle\nonumber \\&=&(X_{b1}\slashed p_{h_i}+X_{b2}\slashed q+X_{b3}m_{h_i})2m_{h_i}X_{b3}\;.
\end{eqnarray}
If $X_{b1}, X_{b2},X_{b3} $ can be expanded as a Taylor series of $q^2 $, each order can be matched. At leading order, $ X_{b2}=0$, or $X_{b2}=X_{b1} $. When $ X_{b2}=0$, $X_{b3}=\pm X_{b1}  $, $ \sum_{spin}	c\left |h_f \right\rangle \left \langle h_f \right |\bar{c}=(\slashed p_{h_f}\pm m_{h_f})X_{b1} $, $ \sum_{spin}	b\left |h_i \right\rangle \left \langle h_i \right |\bar{b}=(\slashed p_{h_i}\pm m_{h_i})X_{b1} $; When $X_{b2}=X_{b1}$, $X_{b3}=\pm X_{b1} \frac{m_{h_i}}{m_{h_f}} $, $ \sum_{spin}	c\left |h_f \right\rangle \left \langle h_f \right |\bar{c}=(\slashed p_{h_i}\pm m_{h_i})X_{b1} $, it should be dropped naturally. The next leading order also faces a similar choice, and for similar reasons all other orders are required to be $ X_{b2}=0,  X_{b3}=\pm X_{b1} $. Then $\sum_{spin}b\left |h_i \right\rangle \left \langle h_i \right |\bar{b}=(\slashed p_{h_i}+m_{h_i})X_b $, it follows naturally $ \left \langle h_i|\bar{b}b| h_i\right\rangle=2m_{h_i}X_b$, the matrix element of initial hadron is
\begin{eqnarray}
	&&\left \langle h_i|\bar{b}\gamma^{\nu}b| h_i\right\rangle =\frac{1}{2}tr[\gamma^{\nu}b_i \left |h_i \right\rangle \left \langle h_i \right |\bar{b}_i]=2p_{h_i}^{\nu}X_{b}\;.
\end{eqnarray}
%另一方面，$ \sum_{spin}(b\left |h_i \right\rangle \left \langle h_i \right |\bar{b})*\left \langle h_i|\bar{b}b| h_i\right\rangle=(X_{b1}\slashed p_{h_i}+X_{b2}\slashed q+X_{b3}m_{h_i})2m_{h_i}X_{b3} $如果$X_{b1}, X_{b2},X_{b3} $可以展开为$q^2 $的泰勒级数，则每一阶都可以匹配。首先是$ X_{b2}=0$或$X_{b2}=X_{b1} $。当$ X_{b2}=0$, $X_{b3}=\pm X_{b1}  $, $ \sum_{spin}	c\left |h_f \right\rangle \left \langle h_f \right |\bar{c}=(\slashed p_{h_f}\pm m_{h_f})X_{b1} $, $ \sum_{spin}	b\left |h_i \right\rangle \left \langle h_i \right |\bar{b}=(\slashed p_{h_i}\pm m_{h_i})X_{b1} $;当$X_{b2}=X_{b1}$、$X_{b3}=\pm X_{b1} \frac{m_{h_i}}{m_{h_f}} $、$ \sum_{spin}	c\left |h_f \right\rangle \left \langle h_f \right |\bar{c}=(\slashed p_{h_i}\pm m_{h_i})X_{b1} $时，应自然删除。下一个阶也面临着类似的选择，出于类似的原因，所有其他阶都被要求为$ X_{b2}=0,  X_{b3}=\pm X_{b1} $ .那么$\sum_{spin}b\left |h_i \right\rangle \left \langle h_i \right |\bar{b}=(\slashed p_{h_i}+m_{h_i})X_b $，自然得出$	\left \langle h_i|\bar{b}b| h_i\right\rangle=2m_{h_i}X_b$，初始强子的矩阵元素为
The maximum value of $\hat{q^2} $ is less than 1 and converges well in most examples. We can do a Taylor series expansion on   $X_{b} (\hat{q^2 })X_{c} (\hat{q^2 })$,
\begin{eqnarray}
	&& X_{b} (\hat{q^2 })X_{c} (\hat{q^2 })=Y_{0}+ Y_{1}\hat{q^2 }+...\;,
\end{eqnarray}
where $  Y_{i} $ are real numbers. It's completely different from the series expansion in LCSR \cite{Bharucha:2015bzk}. So the imaginary part of the matrix element is
\begin{eqnarray}
	&&2 Im M(h_i\to h_i)\nonumber\\
	&	=&\int \frac{d Q^2}{24\pi^2}Im[i\int  \frac{d^{4} q} {(2 \pi)^{4}} \frac{1}{q^{2}-Q^{2}+i \epsilon} \frac{1}{p_{h_f}^{2}-m_{h_f}^{2}+i \epsilon}\nonumber\\
	&&tr[\gamma^{\mu}P_{L}\slashed p_{h_f}\gamma^{\nu}P_{L}\slashed p_{h_i}]( q_{\mu}q_{\nu}-g_{\mu\nu}q^2)\nonumber\\
	&&	(Y_{0}+ Y_{1}\hat{q^2 }+...) ]\;.
\end{eqnarray}

%\begin{widetext}
%	\begin{eqnarray}
%		&&2 Im M(h_i\to h_i)\nonumber\\
%		&	=&\int \frac{d Q^2}{24\pi^2}Im[i\int  \frac{d^{4} q} {(2 \pi)^{4}} \frac{1}{q^{2}-Q^{2}+i \epsilon} %\frac{1}{p_{h_f}^{2}-m_{h_f}^{2}+i \epsilon}tr[\gamma^{\mu}P_{L}\slashed p_{h_f}\gamma^{\nu}P_{L}\slashed p_{h_i}]( %q_{\mu}q_{\nu}-g_{\mu\nu}q^2)(Y_{0}+ Y_{1}\hat{q^2 }+...) ]\;.
%	\end{eqnarray}
%\end{widetext}
%$\hat{q^2} $的最大值小于1，在大多数情况下收敛良好。我们可以对$ X_{b} (\hat{q^2 })$$ X_{b} (\hat{q^2 })=(X_{0}+ X_{1}\hat{q^2 }+...) $做泰勒级数展开，其中$  X_{i} $是实数。它和LCSR中的级数展开完全不同\cite{Bharucha:2015bzk}

The above-proposed hypothesis has a simple physical interpretation. It has a similar physical picture to what Georgi said in the paper\cite{Georgi:1991mr}.
%上述提出的假设有一个简单的物理解释。它有一个类似于作者在论文中所说的物理图片\cite{Georgi:1991mr}。
\begin{eqnarray}
	&&	\left \langle h_i(p_{h_i})|\bar{b}\gamma_{\mu}c|h_f(p_{h_f})\right\rangle\nonumber\nonumber  \\&=&\int \frac{d^3p'_b}{(2\pi) ^3\sqrt{2E'_b}}\bar{u}_{b}(p'_b)\gamma_{\mu}\int \frac{d^3p'_c}{(2\pi) ^3\sqrt{2E'_c}}u_{c}(p'_c)\nonumber\\&&\sqrt{2E_{b}}\sqrt{2E_{c}}\left \langle 0\otimes h_{i}(\text{left})|a_{p_b}a^{\dagger}_{p'_b}a_{p'_c}a^{\dagger}_{p_c}| 0\otimes  h_{f}(\text{left})\right\rangle\nonumber \\
	&=& \left \langle h_{i}(\text{left})|\bar{u}_{b}(p_b)\gamma_{\mu}u_c(p_c) |h_{f}(\text{left})\right\rangle\;.
\end{eqnarray}
When we examine the equations from an alternative perspective, we observe the following transformation:
%当我们换一个角度看等式，我们得到
\begin{eqnarray}
	&&\sum_{spin}c\left |h_f \right\rangle \left \langle h_f \right |\bar{c}\nonumber\\ &\to &\sum_{spin}u_c(p_c)\bar{u}_c(p_c)\left |h_f(\text{left}) \right\rangle \left \langle h_f(\text{left}) \right |\nonumber\\
%	&=&(\slashed p_{c}+m_{c})\left |h_f((\text{left})) \right\rangle \left \langle h_f((\text{left})) \right |\nonumber\\
	&=&(\slashed p_{c}+m_{c})X_c(\hat{q^2 })\;.
\end{eqnarray}
We perform a simultaneous shift, adjusting $X_c(\hat{q^2}) \to X_c(\hat{q^2 })  \frac{m_{h_f}}{m_c} $. This adjustment leads us to the final expressions: %我们同时做一个变换$f(\hat{q^2}) \to f(\hat{q^2 })  \frac{m_{h_f}}{m_c} $。最后的结果是
\begin{eqnarray}
	&\sum_{spin}c\left |h_f \right\rangle \left \langle h_f \right |\bar{c}\to (\slashed p_{h_f}+m_{h_f})X_c(\hat{q^2 }),\nonumber \\&\sum_{spin}b\left |h_i \right\rangle \left \langle h_i \right |\bar{b}\to (\slashed p_{h_i}+m_{h_i})X_b(\hat{q^2 })\;.
\end{eqnarray}
By default, it is assumed that b quarks share the velocity of their respective hadrons, implying that momentum fluctuations should not manifest in the computation of the final decay width.
%默认情况下，b夸克与它们的强子具有相同的速度，并且动量波动不应该出现在最终的衰变宽度中。
\section{Differential decay width under the new theory } 
In the limit that the electron mass is neglected, the differential width of exclusive semi-leptonic decay with respect to the invariant mass of the lepton pair is
%在忽略电子质量的极限下，遍举半轻衰变相对于轻子对不变质量的微分宽度为
\begin{eqnarray}
	\label{q2}
&	&\frac{d \Gamma}{d \hat{q^2 }}*\frac{1}{\Gamma_0}\nonumber\\
%&=& \frac{1}{24\pi^2}Im[i\int  \frac{d^{4} q} {(2 \pi)^{4}} \frac{1}{q^{2}-Q^{2}+i \epsilon} \frac{1}{p_{h_f}^{2}-m_{h_f}^{2}+i \epsilon}\nonumber\\&&tr[\gamma^{\mu}P_{L}\slashed p_{h_f}\gamma^{\nu}P_{L}\slashed p_{h_i}]( q_{\mu}q_{\nu}-g_{\mu\nu}q^2)\nonumber\\&&(Y_{0}+ Y_{1}\hat{q^2 }+...) ]*\frac{8G_{F}^{2}\left|V_{bc}\right|^{2}}{2m_{h_i}}*\frac{m_{h_i}^2}{\Gamma_0}\nonumber\\
	&=&2 \sqrt{\hat{q^4 }-2 (\rho +1) \hat{q^2 }+(\rho -1)^2} \nonumber\\&&\left(-2 \hat{q^4 }+(\rho +1) \hat{q^2 }+(\rho -1)^2\right)(Y_{0}+ Y_{1}\hat{q^2 }+...),\quad		
\end{eqnarray}where $\rho=\frac{m_{h_f}^2}{m_{h_i}^2} ,\Gamma_0=\frac{G_F^2|V_{bc}|^2m_{h_i}^5}{192\pi^3}$.
%The double differential width of exclusive semi-leptonic decay with respect to the  invariant mass of leptons and neutrino energy $ x $ is, 
%与轻子不变质量和中微子能量$ x $有关的遍举半轻衰变的双微分宽度为:
%\begin{eqnarray}
%	\frac{d^2\Gamma}{dx d\hat{q^2} }*\frac{1}{\Gamma_0}	=& 12x (1-x-\rho)(Y_{0}+ Y_{1}\hat{q^2}+...)	,
%\end{eqnarray}
%where $ x=2E_{v_e}/m_{h_i} $. Integrating the double differential width to obtain the single differential width $ \frac{d \Gamma}{d x} $.
%其中$ x=2E_{v_e}/m_{h_i} $。对双微分宽度积分得到单微分宽度$ \frac{d \Gamma}{d x} $。
%\begin{eqnarray}
%	\frac{d \Gamma}{d x}*\frac{1}{\Gamma_0}	&= 	\frac{12x^2 (1-x-\rho)^2}{1-x}[Y_{0}+ \frac{Y_{1}}{2}(\frac{x(1-x-\rho)}{1-x})+...],
%\end{eqnarray}
%The double differential width of exclusive semi-leptonic decay with respect to the invariant mass of leptons and electron energy $ y $ is, 
%与轻子不变质量和中微子能量$ y $有关的遍举半轻衰变的双微分宽度为:
%\begin{eqnarray}
%&&\frac{d^2\Gamma}{dy d\hat{q^2} }*\frac{1}{\Gamma_0}\nonumber\\&=&12 (\hat{q^2}-y)(\rho-1+y-\hat{q^2})(Y_{0}+ Y_{1}\hat{q^2}+...),
%\end{eqnarray}
%where $ y=2E_{e}/m_{h_i} $. Integrating the double differential width to obtain the single differential width $ \frac{d \Gamma}{d y} $
%其中$y=2E_{e}/m_{h_i} $。对双微分宽度积分得到单微分宽度$ \frac{d \Gamma}{d y} $。
%\begin{eqnarray}
%	\frac{d \Gamma}{d y}*\frac{1}{\Gamma_0}	&= 	\frac{2y^2 (\rho +y-1)^2 \left(3 (\rho +1)+2 y^2-(\rho +5) y\right)}{ (1-y)^3}Y_{0}\nonumber\\&-Y_{1}\frac{y^3 (\rho +y-1)^3 \left(4 \rho +y^2-(\rho +3) y+2\right)}{ (y-1)^4}+...,
%\end{eqnarray}
The dot product of initial and final hadrons of four velocities $ \omega =v\cdot v' $ is related to $ \hat{q^2} $ by
%四种速度的初始强子和最终强子的点积$ \omega =v\cdot v' $与$ \hat{q^2} $有关系
\begin{eqnarray}
\omega =\frac{p_{h_i}\cdot p_{h_f}}{m_{h_i}m_{h_f}}=\frac{1+\rho -\hat{q^2}}{2 \sqrt{\rho } }\;.
\end{eqnarray}
The differential width of exclusive semi-leptonic decay with respect to the dot product of initial and final hadrons of four velocities $ \omega =v\cdot v' $ is
%对于四种速度的初始强子和最终强子的点积，遍举半轻衰变的微分宽度为
\begin{eqnarray}
	\frac{d\Gamma}{d \omega}*\frac{1}{\Gamma_0}&=&16 \rho^{3/2}  \sqrt{  \left(\omega ^2-1\right)} \left(3( \rho +1) \omega -2 \sqrt{\rho } \left(2 \omega ^2+1\right)\right)\nonumber\\&&[Y_{0}+ Y_{1}( 1+\rho-2 \sqrt{\rho } \omega )+...]\;.\label{omega1}
\end{eqnarray}	

After integrating the square of the invariant mass of the leptons, we get the total decay width
%在轻子不变质量的平方进行积分后，我们得到了总衰变宽度
\begin{eqnarray}	
	\frac{\Gamma}{\Gamma_0}&=&\left(1-8 \rho+8 \rho^{3}-\rho^{4}-12 \rho^{2} \ln \rho\right)Y_{0}\nonumber\\
	&	&+ Y_{1}[\frac{-3}{10 }(\rho ^5-15 \rho ^4-80 \rho ^3+80 \rho ^2+\nonumber\\&&60 (\rho +1) \rho ^2  \log \rho+15 \rho -1)]+...\;.
\end{eqnarray} Proceed with the calculation of the moments for  $\hat{q^2} $ and $\omega $  to derive the following results:
%计算$q^2 $， x， $\ ω $的矩
\begin{widetext}
\begin{eqnarray}
	\left \langle \hat{q^2} \right \rangle *\frac{\Gamma}{\Gamma_0}\nonumber&=& Y_{0}[\frac{-3}{10 }(\rho ^5-15 \rho ^4-80 \rho ^3+80 \rho ^2+60 (\rho +1) \rho ^2  \log \rho+15 \rho -1)]+\nonumber\\&& Y_{1}[-\frac{2}{15}(\rho ^6-24 \rho ^5-375 \rho ^4+375 \rho ^2+60 \left(3 \rho ^2+8 \rho +3\right) \rho ^2 \log \rho +24 \rho -1)]+...\;,\\
%	\left \langle x \right \rangle *\frac{\Gamma}{\Gamma_0}&= &Y_{0}(\frac{2 \rho ^5}{5}-3 \rho ^4+12 \rho ^3-4 \rho ^2-12 \rho ^2 \log \rho-6 \rho +\frac{3}{5})+\nonumber \\&&Y_{1}[\frac{\rho ^6}{10}-\frac{6 \rho ^5}{5}+9 \rho ^4+24 \rho ^3-\frac{57 \rho ^2}{2}-18 \rho ^2 \log \rho+\rho  \left(-24 \rho ^2 \log \rho-\frac{18}{5}\right)+\frac{1}{5}]+...,	\\	
%	\left \langle y \right \rangle *\frac{\Gamma}{\Gamma_0}&= &\frac{Y_{0}}{10}(7-75 \rho -120 \rho ^2+200 \rho ^3-15 \rho ^4+3 \rho ^5-  60\rho ^2(\rho +3) \log\rho )+\frac{Y_{1}}{30} \nonumber\\&& (2 \rho ^6-12 \rho ^5 +645 \rho ^4-180 \rho ^4 \log\rho +720 \rho ^3-1200\rho ^3\log \rho   -1230 \rho ^2-720 \rho ^2 \log \rho -132 \rho +7)+...,\quad\\
	\left \langle \omega \right \rangle *\frac{\Gamma}{\Gamma_0}&=& Y_{0}[3 (\rho +1) \rho ^{3/2}   \log\rho +\frac{1}{20}\sqrt{\rho +\frac{1}{\rho }-2}(7 \rho ^4-18 \rho ^3+ 142 \rho ^2-18 \rho +7)]\nonumber \\&&+Y_{1} [\left(3 \rho ^2+14 \rho +3\right) \rho ^{3/2}  \log \rho+\frac{1}{12}\sqrt{\rho +\frac{1}{\rho }-2}  \left(\rho ^5-5 \rho ^4+124 \rho ^3+124 \rho ^2-5 \rho +1\right) ]+...\;,\quad\quad
\end{eqnarray} 
\end{widetext}
 If we assume that the corresponding contribution always decreases as the order $ \hat{q^2 } $ increases. Similarly, in the heavy quark effective theory, there exists a Taylor series expansion at $\Lambda_{qcd}/M_B$, which also assumes that the higher the series, the lower the contribution.  To facilitate the fitting of experimental data, we can truncate the total decay width and moment from the theory. If we want to solve for m unknown numbers ($Y_{0},... Y_{m-1}$) , other parameters ($Y_{m},Y_{m+1},...$) can be set to 0, m equations are listed from the total decay width and moment. The physical basis for this approach lies in the fact that the higher-order terms correspond to contributions further away from the boundary of the physical interval, and their values are suppressed by both the phase space and the strong interaction shape factor; thus, higher-order contributions are expected to be suppressed. Therefore, under the current experimental accuracy, it is sufficient to retain only the first two terms, and the omission of higher-order terms will not introduce considerable bias.
%如果我们假设相应的贡献总是随着阶次$ \hat{q^2 } $的增加而减小。因此，我们可以从理论中截断总衰减宽度和总衰减矩，从实验数据中得到总衰减宽度和总衰减矩。如果我们想求解M个未知数($ Y_{0},... Y_{m-1}，就可以令其他参数($ Y_{m},Y_{m+1}...$)为0，从总衰变宽度和矩中列出M个方程。 $)。
%$\hat{q^2} $的最大值小于1，在大多数情况下收敛良好。我们可以对$ X_{b} (\hat{q^2 })$$ X_{b} (\hat{q^2 })=(X_{0}+ X_{1}\hat{q^2 }+...) $做泰勒级数展开，其中$  X_{i} $是实数。它和LCSR中的级数展开完全不同\cite{Bharucha:2015bzk}所以总衰减宽度是

The previous calculation is considered to be a strict V-A current; next, we think the coefficients can be different for vector and axial vector currents.%前面的计算被认为是严格的V-A流,接下来我们认为矢量流和轴向流的系数可以不同。
\begin{eqnarray}
	V : \quad		&\sum_{spin}c\left |h_f \right\rangle \left \langle h_f \right |\bar{c}\to (\slashed p_{h_f}+m_{h_f})f_c(\hat{q^2 }) ,\nonumber\\&\sum_{spin}b\left |h_i \right\rangle \left \langle h_i \right |\bar{b}\to (\slashed p_{h_i}+m_{h_i})f_b(\hat{q^2 })\;,\\
	A: \quad		&\sum_{spin}c\left |h_f \right\rangle \left \langle h_f \right |\bar{c}\to (\slashed p_{h_f}+m_{h_f})g_c(\hat{q^2 }) ,\nonumber\\&\sum_{spin}b\left |h_i \right\rangle \left \langle h_i \right |\bar{b}\to (\slashed p_{h_i}+m_{h_i})g_b(\hat{q^2 })\;.
\end{eqnarray} 
 The differential width with respect to the invariant mass of leptons is, 
%关于轻子不变质量的微分宽度是
\begin{eqnarray}	
	&&\frac{d \Gamma}{d \hat{q^2 }}*\frac{1}{\Gamma_0}\nonumber\\ %&=&\frac{G_{F}^{2}\left|V_{bc}\right|^{2}m_{h_i}}{6\pi^2\Gamma_0}Im[i\int  \frac{d^{4} q} {(2 \pi)^{4}} \frac{1}{q^{2}-Q^{2}+i \epsilon} \frac{1}{p_{h_f}^{2}-m_{h_f}^{2}+i \epsilon}\nonumber\\
	%&&tr[\gamma^{\mu}\frac{f-g\gamma_{5}}{2}\slashed p_{h_f}\gamma^{\nu}\frac{f-g\gamma_{5}}{2}\slashed p_{h_i}]( q_{\mu}q_{\nu}-g_{\mu\nu}q^2) ]\nonumber\\
	&=&2 \sqrt{\hat{q^4 }-2 (\rho +1) \hat{q^2 }+(\rho -1)^2} [\left(f^2-g^2\right)6\sqrt{\rho }\hat{q^2 }\nonumber\\&&+(f^2+g^2)\left(-2 \hat{q^4 }+(\rho +1) \hat{q^2 }+(\rho -1)^2\right)]\;,	
\end{eqnarray} 
where $ f=\sqrt{f_bf_c}$, $g=\sqrt{g_bg_c} $. If $ f(\hat{q^2 }),g(\hat{q^2 }) $ keep only the first constant term, the total differential width is
%如果$ f(\hat{q^2}) g(\hat{q^2}) $只保留第一个常数项，总微分宽度为
\begin{eqnarray}	
	\frac{\Gamma}{\Gamma_0}&=&\frac{\left(f^2+g^2\right)}{2}\left(1-8 \rho+8 \rho^{3}-\rho^{4}-12 \rho^{2} \log \rho\right)+ (f^2 \nonumber\\&& -g^2)\sqrt{\rho } \left(\rho ^3+9 \rho ^2-9 \rho -6 (\rho +1) \rho  \log \rho-1\right)\;.
\end{eqnarray}  

\section{baryon } 
%自旋1/2到自旋1/2的最一般的重子衰变形状因子可以写成这种形式。
The most general baryon decay form factors for spin 1/2 to spin 1/2 can be written in the form:
\begin{eqnarray}
	\label{baryon}
	&& \left \langle \mathcal{B}_c(p_{h_f},s')|\bar{c}\gamma^{\mu} b| \mathcal{B}_b(p_{h_i},s)\right\rangle \nonumber\\ &=&\bar{u}(p_{h_f},s')[f_1\gamma^{\mu}+if_2\sigma^{\mu\nu}q_{\nu}+f_3q^{\mu}]u(p_{h_i},s)\;,\\
	&&	\left \langle \mathcal{B}_c(p_{h_f},s')|\bar{c}\gamma^{\mu}\gamma^{5}b| \mathcal{B}_b(p_{h_i},s)\right\rangle \nonumber\\ &=&\bar{u}(p_{h_f},s')[g_1\gamma^{\mu}+ig_2\sigma^{\mu\nu}q_{\nu}+g_3q^{\mu}]\gamma^{5} u(p_{h_i},s)\;,
\end{eqnarray}
where $ q=p_{h_i}-p_{h_f} $, $\sigma^{\mu\nu}=i[\gamma^{\mu},\gamma^{\nu}]/2  $,   $f_i,g_i$ are the scalar function of $ \hat{q^2}$, $\mathcal{B}_b  $ is a baryon contain a bottom quark. Since quarks in baryons have the same velocity as baryons, then the spinor of baryons differs from the spinor of quarks by only one proportional factor,
\begin{eqnarray}
	u(p_{h_i},s)=\sqrt{\frac{m_{h_i}}{m_b}}u(p_{b},s)\;,
\end{eqnarray}adjusting $f_i \to \sqrt{\frac{m_b}{m_{h_i}}}f_i$\;, $g_i \to \sqrt{\frac{m_b}{m_{h_i}}}g_i $\;,\;so we can use quark spinors instead of baryon spinors.
%其中$ q=p_{h_i}-p_{h_f} $, $\sigma^{\mu\nu}=i[\gamma^{\mu},\gamma^{\nu}]/2  $, $f_i,g_i$是$ \hat{q^2}$的函数，$\mathcal{B}_b  $是包含一个b夸克的重子。由于重子中的夸克具有与重子相同的速度，那么重子的自旋量与夸克的自旋量只相差一个比例因子，所以我们可以用夸克自旋来代替重子自旋。
\begin{eqnarray}
&	&\left \langle \mathcal{B}_c(p_{h_f},s')|\bar{c}\gamma^{\mu} b| \mathcal{B}_b(p_{h_i},s)\right\rangle \nonumber\\ &=&\bar{u}(p_{c},s')[f_1\gamma^{\mu}+if_2\sigma^{\mu\nu}q_{\nu}+f_3q^{\nu}]u(p_{b},s)\;, \\
&	&\left \langle \mathcal{B}_c(p_{h_f},s')|\bar{c}\gamma^{\mu}\gamma^{5}b| \mathcal{B}_b(p_{h_i},s)\right\rangle \nonumber\\ &=&\bar{u}(p_{c},s')[g_1\gamma^{\mu}+ig_2\sigma^{\mu\nu}q_{\nu}+g_3q^{\nu}]\gamma^{5} u(p_{b},s)\;.
\end{eqnarray}

\begin{figure}[!ht]
	\centering
	\includegraphics[width=0.8\linewidth]{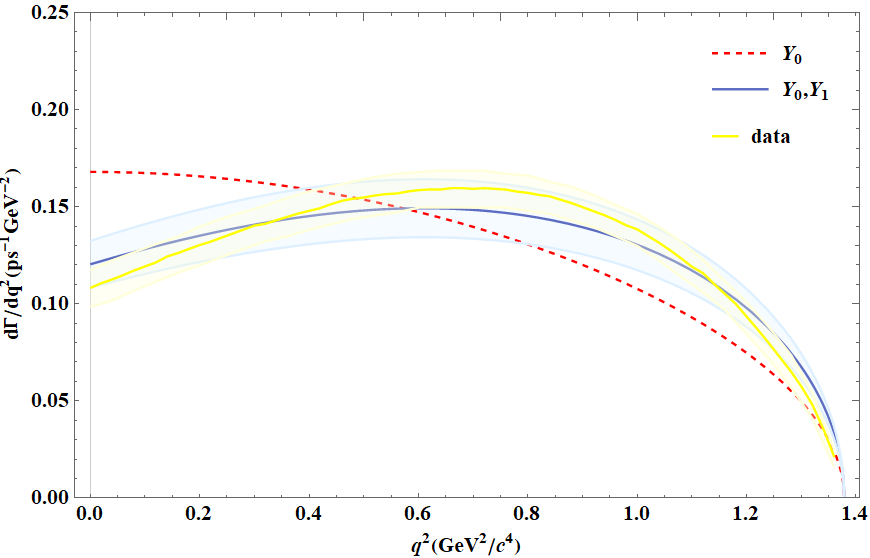}
	\caption{Comparison of the differential decay rates of $\Lambda_c \to \Lambda+e^{+}+v_{e}$ \cite{BESIII:2022ysa}with different parameters. The blue line has two parameters  with total uncertainty,  the red line has one parameter, and the yellow dots are data point.}
		\label{fig:1}
\end{figure}

When we consider all the parameters $ f_1$,$f_2$,$f_3$, $g_1$,$g_2$,$g_3 $, the amplitude of the hadronic component in hadronic semi-leptonic decay is
%当我们考虑所有参数$ f_1$，$f_2$，$f_3$， $g_1$，$g_2$，$g_3 $时，强子半轻衰变中强子分量的振幅为
\begin{eqnarray}
	&&\mathcal{M}^{\mu}\mathcal{M}^{\nu}\nonumber\\&=&Tr[((f_1-g_1\gamma_{5})\gamma^{\mu}+i(f_2-g_2\gamma_{5})\sigma^{\mu\nu}q_{\nu}+(f_3-g_3\gamma_{5})q^{\mu})\nonumber\\&& (\slashed{p_{h_i}}+m_{h_i})((f_1-g_1\gamma_{5})\gamma^{\nu}+i(f_2-g_2\gamma_{5})\sigma^{\nu\lambda}q_{\lambda}\nonumber\\&&+(f_3-g_3\gamma_{5})q^{\nu})(\slashed{p_{h_f}}+m_{h_f})]\;.
\end{eqnarray} 
We defined the integrated leptonic tensor as
%我们将积分轻子张量定义为
\begin{eqnarray}	
	\mathcal{L}^{\mu\nu}
	%&=&\int[d^3p_e][d^3p_{\nu_e}]L^{\mu\nu}(p_e,p_{\nu_e})(2\pi)^4\delta^4(q-p_e-p_{\nu_e})\nonumber\\
	&=&\frac{\left(q^2-m_l^2\right)^2 }{24 \pi  q^6}\left[2 q^{\mu } q^{\nu } \left(2 m_l^2+q^2\right)\right.\nonumber\\
	&&\left.-q^2 \left(m_l^2+2 q^2\right) g^{\mu  \nu }\right]\;.
\end{eqnarray} 
%where $  \sigma=\frac{m_l^2}{m_{h_i}^2}$. 
%The integrated full amplitude is，
%其中$  \sigma=\frac{m_l^2}{m_{h_i}^2}$。积分过的全部振幅是，
%\begin{widetext}
%\begin{eqnarray}	
%	&&\mathcal{M}_{\mu}\mathcal{M}_{\nu}\mathcal{L}^{\mu\nu}\nonumber\\&=& (\left(f_1^2+g_1^2\right) (\hat{q}^4 (\rho -\sigma +1)+\hat{q^2} \left((\rho -1)^2-(\rho +1) \sigma \right)+2 (\rho -1)^2 \sigma -2 \hat{q}^6)+6 \sqrt{\rho } \hat{q}^4 \left(g_1^2-f_1^2\right))\frac{m_{\text{h_i}}^4 \left(\sigma -\hat{q^2}\right)^2}{6 \pi  \hat{q}^6}\nonumber\\&&+\frac{\sigma  m_{\text{h_i}}^6 \left(\sigma -\hat{q^2}\right)^2 \left(\left(f_3^2-g_3^2\right) \left(-\hat{q^2}+\rho +1\right)+2 \sqrt{\rho } \left(f_3^2+g_3^2\right)\right)}{4 \pi  \hat{q^2}}\nonumber\\&&+
%	\frac{m_{\text{h_i}}^6 \left(\sigma -\hat{q^2}\right)^2 \left(2 \hat{q^2}+\sigma \right) \left(6 \sqrt{\rho } \hat{q^2} \left(f_2^2+g_2^2\right)+\left(f_2^2-g_2^2\right) \left((\rho +1) \hat{q^2}+\hat{q}^4-2 (\rho -1)^2\right)\right)}{12 \pi  \hat{q}^4}\nonumber\\&&-\frac{\left(\sqrt{\rho }-1\right) m_{\text{h_i}}^5 \left(\left(\sqrt{\rho }+1\right)^2-\hat{q^2}\right) \left(\sigma -\hat{q^2}\right)^2 \left(f_1 f_3 \sigma -g_1 g_2 \left(2 \hat{q^2}+\sigma \right)\right)}{2 \pi\hat{q}^4}.\quad
%\end{eqnarray} 
%\end{widetext}
In the limit that the electron mass is neglected and contributions from $ f_2,g_2,f_3 $ and $ g_3 $ are concurrently disregarded, the modulus squared of the decay amplitude is reduced to a simplified form:
%m_l→0$的极限中，我们同时忽略f_2,g_2,f_3,g_3 $，衰变振幅模方化简为
\begin{eqnarray}	
	&&\mathcal{M}_{\mu}\mathcal{M}_{\nu}\mathcal{L}^{\mu\nu}\nonumber\\
	&=&	[(\hat{q^2} (\rho +1)+ (\rho -1)^2-2 \hat{q}^4)\left(f_1^2+g_1^2\right) \nonumber\\
	&&+6 \sqrt{\rho } \hat{q^2} \left(f_1^2-g_1^2\right)]\frac{ m_{h_i}^4}{6\pi }\;.
\end{eqnarray}
We select the differential width  $ \frac{d\Gamma}{d q^2} $ of $ \Lambda_c^{+}\to  \Lambda +e^{+}+v_{e} $ in the article \cite{BESIII:2022ysa},  the moment of $ q^2 $ is obtained by the experimental data, and then the parameters $ Y_i $ are obtained by fitting eq (\ref{q2}), and with the increase of the number of fitting parameters, the fitting effect is better. We found that the fitting to the second parameter in FIG.~\ref{fig:1} is already very close to the experimental curve, and the results for the parameters are 
\begin{eqnarray}	
	&Y_0= 0.328\pm0.010, Y_1= 1.184\pm0.092.
\end{eqnarray}
The fitting to the higher order is of little significance, because the actual experiment must only get data points, and there is already a large error in the fitting curve.
%我们选择 在文章中 \cite{BESIII:2022ysa}$ \Lambda_c^{+}\to  \Lambda +e^{+}+v_{e} $的微分宽度  $ \frac{d\Gamma}{d q^2} $ $ q^2 $ 的矩 是由实验数据得到的，然后是参数 $ Y_i $ 由公式(13)拟合得到，随着拟合参数个数的增加，拟合效果越好。我们发现对第二个参数的拟合已经非常接近实验曲线，对高阶的拟合意义不大，因为实际实验必须只得到数据点，而拟合曲线已经存在较大的误差。
\begin{figure}[!ht]
	\centering
	\includegraphics[width=0.8\linewidth]{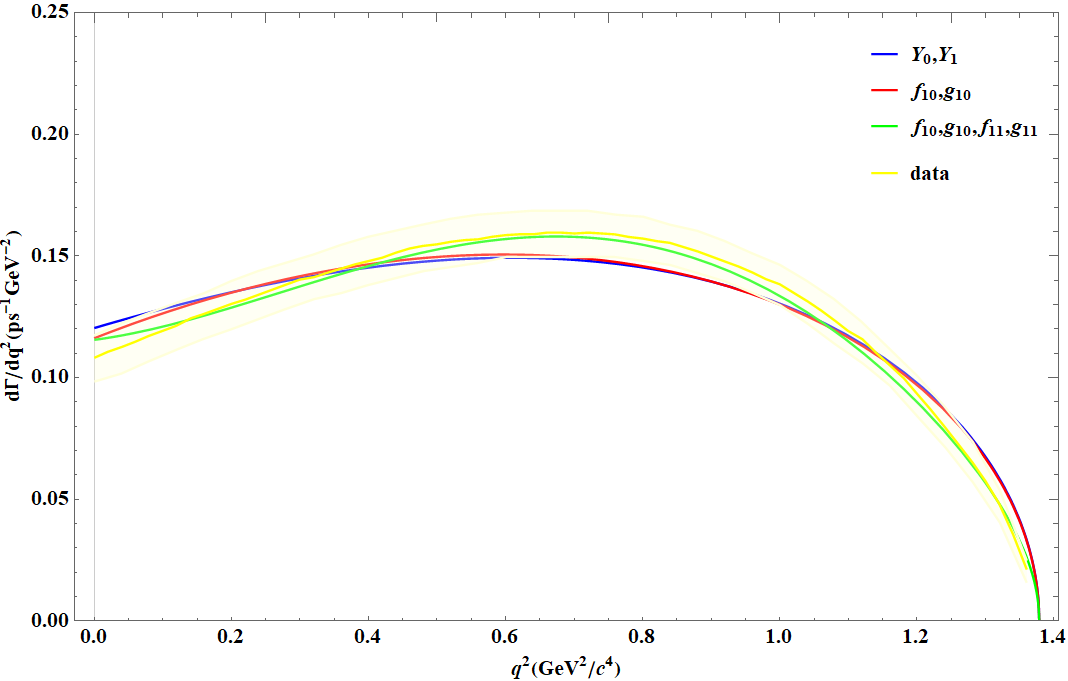}
	\caption{Comparison of the differential decay rates of $\Lambda_c \to \Lambda+e^{+}+v_{e}$ \cite{BESIII:2022ysa}with different parameters. The blue line has two parameters, the red line has two parameters,  the green line has four parameters , and the yellow dots are data point. }	\label{fig:3}
\end{figure}\\
We also fit in the case where $ f_1 \neq g_1$ to get 2 and 4 parameters in FIG.~\ref{fig:3}, where $f_{10},g_{10}$  represents the first constant term of $f_1,g_1$ taylor expansion, $f_{11},g_{11}$ represents the secong term of $f_1,g_1$ taylor expansion. They are very similar to the case of 2 parameters satisfying V-A. Considering the large error in the experimental data, it is not clear whether V-A is satisfied.
%我们也在这种情况下拟合得到2参数和4参数在图像3。他们跟满足V-A的2参数情况是非常相似的，考虑到实验数据存在较大的误差，目前还不能明确区分是否满足V-A。

If we directly expand the Taylor series for the differential width $\frac{d \Gamma}{d\hat{q^2}}  $, which is
%我们直接展开泰勒级数求微分宽度$\frac{d \Gamma}{d\hat{q^2}}  $，也就是
\begin{eqnarray}
	\frac{d \Gamma}{d\hat{q^2}}*\frac{1}{\Gamma_0}=&Y_{0}+ Y_{1}\hat{q^2}+ Y_{2}\hat{q^4}+...\;.
\end{eqnarray}
The total decay width and moment are respectively%总衰减宽度和矩分别为
\begin{eqnarray}	
	\frac{ \Gamma}{\Gamma_0}=&(1-\sqrt{\rho})^2Y_{0}+ \frac{(1-\sqrt{\rho})^4}{2}Y_{1}+\frac{(1-\sqrt{\rho})^6}{3}Y_{2}+...,\;\nonumber\\
	  \frac{\left \langle \hat{q^2} \right \rangle\Gamma}{\Gamma_0}=&\frac{(1-\sqrt{\rho})^4}{2}Y_{0}+\frac{(1-\sqrt{\rho})^6}{3}Y_{1}+\frac{(1-\sqrt{\rho})^8}{4}Y_{2}+...\;\quad
\end{eqnarray}

If we still assume that the contribution of the higher order is much smaller than that of the lower order, the result is very poor. It is impossible to fit the results close to the experimental curve.
%如果我们仍然假设高阶的贡献比低阶的贡献小得多，结果就会很差。要使结果接近实验曲线是不可能的。
\section{meson} 
In Heavy Quark Physics  \cite{Manohar:2000dt} section 2.9 or Elementary Particles and Their Interaction Concepts and Phenomena  \cite{Quang:1998yw} section 10.1, the most common matrix element of $ \bar{B}\to D $ vector current and axial current  is
%在重夸克物理2.9节\cite{Manohar:2000dt}或基本粒子及其相互作用概念和现象10.1节\cite{Quang:1998yw}中，$ \bar{B}\to D $矢量流和轴矢量流最常见的矩阵元素是
\begin{eqnarray}
	\left\langle D\left(p^{\prime}\right)\left|V^\mu\right|\bar{B}(p)\right\rangle&=&  f_{+}\left(q^2\right)\left(p+p^{\prime}\right)^\mu+f_{-}\left(q^2\right)\left(p-p^{\prime}\right)^\mu\;, \nonumber\\
	\left\langle D\left(p^{\prime}\right)\left|A^\mu\right|\bar{B}(p)\right\rangle &=&0\;,\label{Vector}
\end{eqnarray}
where $q=p-p^{\prime}$, all the form factors are real, and the states have the usual relativistic normalization. This situation seems to be different from the baryon case and incompatible with our scheme. If the current operator is composed of field operators of hadrons, this is no problem, but now the field operator in the current operator and the state are not at the same level, and then this parameterization is questionable. We're going to go through a series of proofs that there are two kinds of vectors. One consists of four momenta , and the other consists of bilinears. The vector current operators $ \bar{b}\gamma^{\mu}b, \bar{b}\gamma^{\mu}c $ and axial vector current operators $ \bar{b}\gamma^{\mu}\gamma^{5}b, \bar{b}\gamma^{\mu}\gamma^{5}c $ are Lorentz vectors because of the spinor field change factor is only related to velocity, and its matrix elements $\left \langle T_b(p',s')|\bar{b}\gamma^{\mu}b| T_b(p,s)\right\rangle  $, $\left \langle T_b(p',s')|\bar{b}\gamma^{\mu}c| T_c(p,s)\right\rangle  $, $\left \langle T_b(p',s')|\bar{b}\gamma^{\mu}\gamma^{5}b| T_b(p,s)\right\rangle  $ and $\left \langle T_b(p',s')|\bar{b}\gamma^{\mu}\gamma^{5}c| T_c(p,s)\right\rangle  $  are also Lorentz vectors, where $\left| T_b(p,s)\right\rangle $ is a particle state with momentum p and spin s containing b quarks, it could be a quark, a meson or a baryon. For quark states
%其中$q=p-p^{\prime}$，所有的形式因子都是实数，并且状态具有通常的相对论规一化。这种情况似乎与重子的情况不同，与我们的方案不相容。如果流算符是由强子的场算符组成的，这是没有问题的，但是现在流算符中的场算符和态不在同一个层次上，那么这种参数化是有问题的。我们将通过一系列的证明来证明有两种向量。一个由四个动量组成，另一个由两个旋量组成。矢量流算符$ \bar{b}\gamma^{\mu}b, \bar{b}\gamma^{\mu}c $和轴向矢量流算符$ \bar{b}\gamma^{\mu}\gamma^{5}b, \bar{b}\gamma^{\mu}\gamma^{5}c $是洛伦兹矢量，因为自旋场变化因子只与速度有关，其矩阵元$\left \langle T_b(p',s')|\bar{b}\gamma^{\mu}b| T_b(p,s)\right\rangle  $、$\left \langle T_b(p',s')|\bar{b}\gamma^{\mu}c| T_c(p,s)\right\rangle  $、$\left \langle T_b(p',s')|\bar{b}\gamma^{\mu}\gamma^{5}b| T_b(p,s)\right\rangle  $和$\left \langle T_b(p',s')|\bar{b}\gamma^{\mu}\gamma^{5}c| T_c(p,s)\right\rangle  $也是洛伦兹矢量，其中$\left| T_b(p,s)\right\rangle $是动量为p、自旋为s的粒子态，包含b个夸克，它可以是夸克、介子或重子。对于夸克态
\begin{eqnarray}
	&\left \langle b(p,s)|\bar{b}\gamma^{\mu}b| b(p,s)\right\rangle=\bar{u}_{b}(p,s)\gamma^{\mu}u_{b }(p,s)=2p^{\mu}\;,
\end{eqnarray}for hardon states
\begin{eqnarray}
	&\left \langle H_b(p,s)|\bar{b}\gamma^{\mu}b| H_b(p,s)\right\rangle\propto p^{\mu}\;.
\end{eqnarray}
Different people have different opinions about the former coefficient. In the heavy quark effective theory of inclusive decay, the coefficient is 2. It is considered that the coefficients obtained by considering the conservation of charge are consistent with those that occur in real decay. But I think it's two different things, the coefficient obtained by considering the conservation of charge is 2, because it doesn't account for the strong interaction. In the limit where the mass of the heavy quark goes to infinity, this coefficient goes to 2. The matrix element should take into account the final particles in the real decay , the coefficient should be the scalar function of $ \hat{q^2} $. And it's also true for the decay of light quarks.  For $ p\not =p' $, there should be a question for $\left \langle T_b(p',s)|\bar{b}\gamma^{\mu}b| T_b(p,s)\right\rangle =a p^{\mu}+ b p'^{\mu} $, $\left \langle T_b(p',s)|\bar{b}\gamma^{\mu}c | T_c(p,s)\right\rangle =a p^{\mu}+ b p'^{\mu} $, where $ a,b $ are lorentz scalar. We will take the same particle with different velocities to show that such a,b does not exist, and likewise the case of different particles with different velocities.
%不同的人对前一个系数有不同的看法。在单举衰变的重夸克有效理论中，该系数为2。它们认为考虑电荷守恒所得到的系数与实际衰变中的系数是一致的。但我认为这是两个不同的东西，考虑电荷守恒得到的系数是2，因为它没有考虑强相互作用。当重夸克的质量趋于无穷时，这个系数趋于2。矩阵元应考虑到最终粒子在实际衰变中，系数应为$ \hat{q^2} $的标量函数。轻夸克的衰变也是如此。对于$ p\not =p' $，应该有一个关于$\left \langle T_b(p',s)|\bar{b}\gamma^{\mu}b| T_b(p,s)\right\rangle =a p^{\mu}+ b p'^{\mu} $$\left \langle T_b(p',s)|\bar{b}\gamma^{\mu}c | T_c(p,s)\right\rangle =a p^{\mu}+ b p'^{\mu} $的问题，其中$ a,b $是洛伦兹标量。我们将用不同速度的相同粒子来证明这样的a,b不存在，同样，不同速度的不同粒子的情况也是如此。
For the quark state, we list the Gordon identity%对于夸克态，我们列出戈登恒等式
\begin{eqnarray}
	&&\left \langle b(p',s')|\bar{b}\gamma^{\mu}b| b(p,s)\right\rangle=\bar{u_b}(p^{\prime},s')\gamma^{\mu}u_b(p,s)\nonumber\\&=&\bar{u_b}(p^{\prime},s')\left[\frac{p^{\prime\mu}+p^{\mu}}{2m}+\frac{i\sigma^{\mu\nu}q_{\nu}}{2m}\right]u_b(p,s)\;.
\end{eqnarray}
we generalize the Gordon identity to different masses, %对于重子态，我们将戈登恒等式推广到不同质量
\begin{eqnarray}
	&&\left \langle b(p',s')|\bar{b}\gamma^{\mu}c| c(p,s)\right\rangle=\bar{u_b}(p^{\prime},s')\gamma^{\mu}u_c(p,s)\nonumber\\&=&\bar{u_b}(p^{\prime},s')\left[\frac{p^{\prime\mu}+p^{\mu}}{m_b+m_c}+\frac{i\sigma^{\mu\nu}q_{\nu}}{m_b+m_c}\right]u_c(p,s)\;.
\end{eqnarray}
For the baryon state, we can rewrite the eq (\ref{baryon}) as a four-momentum combination structure.%我们可以将eq (\ref{baryon})重写为一个四动量组合结构。
\begin{eqnarray}	
&	&\left \langle \mathcal{B}_c(p_{h_f},s')|\bar{c}\gamma^{\mu} b| \mathcal{B}_b(p_{h_i},s)\right\rangle \nonumber\\&=&\bar{u}_{\mathcal{B}_c}[\frac{f_1}{(m_{h_i}+m_{h_f})}(p^{\mu}_{h_i}+p_{h_f}^{\mu})\nonumber\\
&	&+(\frac{f_1}{m_{h_i}+m_{h_f}}+f_2)i\sigma^{\mu\nu}q_{\nu}+f_3q^{\mu}]u_{\mathcal{B}_b }\;.	
\end{eqnarray} 
 
My point is that meson matrix elements can also be parameterized just as baryons do. The meson matrix element from a pseudoscalar particle to a vector particle is exactly the same as the meson matrix element from a pseudoscalar particle to a pseudoscalar particle, but the scalar function is different. Since the matrix elements do not exist alone, it is necessary to sum the quark spins at the same time as the module squares. 
%我的观点是介子矩阵元素也可以像重子一样参数化。从赝标量粒子到矢量粒子的介子矩阵元素与从赝标量粒子到赝标量粒子的介子矩阵元素完全相同，只是标量函数不同。由于矩阵元素不是单独存在的，因此有必要在模块平方的同时对夸克自旋求和。
\begin{eqnarray}
	&&\left\langle D^{(*)}\left(p^{\prime}\right),s'\left|V^\mu\right|\bar{B}(p),s\right\rangle\nonumber\\ &\to &\bar{u}(p_{c},s')[f_1\gamma^{\mu}+if_2\sigma^{\mu\nu}q_{\nu}+f_3q^{\nu}]u(p_{b},s)\;, \label{meson matrix1}\\
	&&\left\langle D^{(*)}\left(p^{\prime}\right),s'\left|A^\mu\right|\bar{B}(p),s\right\rangle \nonumber\\ &\to &\bar{u}(p_{c},s')[g_1\gamma^{\mu}+ig_2\sigma^{\mu\nu}q_{\nu}+g_3q^{\nu}]\gamma_{5}u(p_{b},s)\;.\label{meson matrix2}
\end{eqnarray}
The parametric form (Equations (\ref{meson matrix1})-(\ref{meson matrix2})) adopted for the meson matrix elements in this section is a working assumption based on unity inspiration, and its correctness still needs to be verified by future precise experimental data. Therefore, this part of the framework should be regarded as an exploratory result.
%本节对介子矩阵元所采用的参数化形式（式(37)-(44)）系基于统一性启发提出的工作假设，其正确性仍需由未来精密实验数据予以检验；因此，该部分框架应被视为探索性结果。

So, the axial vector from B meson to B meson is%所以，从B介子到B介子的轴矢量是
\begin{eqnarray}
	&\left\langle \bar{B}(p)\left|A^\mu\right| \bar{B}(p)\right\rangle  =\frac{g_1}{2}\sum_s\bar{u}(p_{b},s)\gamma^{\mu}\gamma_{5}u(p_{b},s)=0\;.\quad\quad
\end{eqnarray}

\begin{figure}[!ht]
	\centering
	\includegraphics[width=0.8\linewidth]{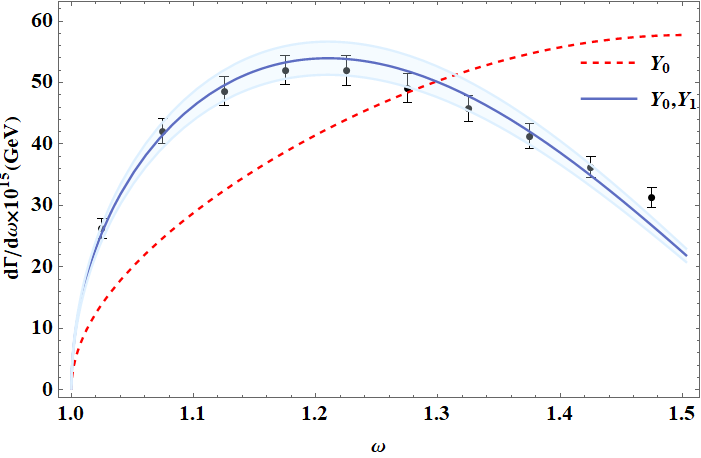}
	\caption{Comparison of the differential decay rates of $\bar{B}^0\to  D^{*+} +e^{-}+v_{e}$\cite{Belle-II:2023okj} with different parameters. The blue line has two parameters with total uncertainty, and the red line has one parameter, and the black dots are data with total uncertainty.}
	\label{fig:2}
\end{figure}

In the limit that the electron mass is neglected, $ g_1=f_1 $ and contributions from $ f_2,g_2,f_3 $ and $ g_3 $ are concurrently disregarded. 
After the spin summation, the hadron tensor $h_{\mu\nu}$ becomes
$\frac{f_1^2}{4}tr[\gamma^{\mu}P_{L}\slashed p_{c}\gamma^{\nu}P_{L}\slashed p_{b}]$,
therefore, \begin{eqnarray}
	&\frac{f_1^2m_{c}m_{b}}{4m_{h_i}m_{h_f}}=X_{b} (\hat{q^2 })X_{c} (\hat{q^2 }).
\end{eqnarray}When we calculate the differential width $ \frac{d\Gamma}{d \omega} $ of $ \bar{B}^0\to  D^{*+} +e^{-}+v_{e} $ in the article \cite{Belle-II:2023okj}, we directly use  the results containing $X_{b} (\hat{q^2 })X_{c} (\hat{q^2 })$, the moment of $\omega$ is obtained by the experimental data, and then the parameters $ Y_i $ are obtained by fitting eq. (\ref{omega1}). 
The results for the parameters are 
\begin{eqnarray}	
	&Y_0= 0.152\pm0.008, Y_1= 1.606\pm0.080.
\end{eqnarray}
As displayed in FIG.~\ref{fig:2}, with the increase of the number of fitting parameters, the fitting effect is better. 
%我们选择微分宽度  $ \frac{d\Gamma}{d \omega} $ 的 $ \bar{B}\to  D^* +e^{+}+v_{e} $ 在文章中 \cite{BelleII:2023lst}的时刻 $\omega$ 是由实验数据得到的，然后是参数 $ Y_i $ 由公式()拟合得到。就像2展示的，随着拟合参数个数的增加，拟合效果越好。

\section{the decay of pseudoscalar mesons into pseudoscalar mesons}
%在忽略轻子质量的极限下，根据已有定义，可得$\bar{B}\to De\bar{\nu}_e$不变矩阵元为
 According to the eq.(\ref{Vector}), in the limit where the lepton mass is neglected,
the matrix element for $\bar{B} \to D e \bar{\nu}_e$ is 
\begin{eqnarray}
	&&\mathcal{M}(\bar{B}\rightarrow De\bar{v}_{e})\nonumber\\&=&\sqrt{2}G_{F}V_{cb}\:f_{+}(p+p^{\prime})^{\mu}\:\bar{u}(p_{e})\gamma_{\mu}P_{L}v\big(p_{\nu_{e}}\big).
\end{eqnarray}
%该矩阵元对二阶强子张量有贡献的只有W_2。由于W_3=0，电子相关的双微分宽度$\frac{d^2 \Gamma}{d \hat{q^2} dy}*\frac{1}{\Gamma_0}$和电子型中微子相关的双微分宽度$\frac{d^2 \Gamma}{d \hat{q^2} dx}*\frac{1}{\Gamma_0}$表达式是一致的，
the matrix element contributes to the hadronic tensor only through $ W_2 $. Due to $ W_3 = 0 $, the expressions for the double differential widths related to electrons $ \frac{d^2 \Gamma}{d \hat{q^2} dy} * \frac{1}{\Gamma_0} $, and those related to electron neutrinos $ \frac{d^2 \Gamma}{d \hat{q^2} dx} * \frac{1}{\Gamma_0} $ are same,
\begin{eqnarray}
	\frac{d^2 \Gamma}{d \hat{q^2} dz}*\frac{1}{\Gamma_0}=6|f_{+}|^{2} [\hat{q^2} (z-1)-z (\rho +z-1)]\;,		
\end{eqnarray}
 where $ y,x $ represents the dimensionless energy of the electron and electron neutrino, $y=\frac{2E_e}{m_{hi}},x=\frac{2E_{v_e}}{m_{hi}}$, where $ z $ represents $ y $ or $x $.

%其中$z$为无量纲化的电子或者电子型中微子的能量。

%我的理论，($  B \to D$)f和g近似相等，在$ \hat{Q}^2$一个狭窄条范围内，单个轻子能量的分布与自由夸克层次是相同的，只差一个形状因子。\\

In the limit that the electron mass is neglected and contributions from $ f_2,g_2,f_3 $ and $ g_3 $ are concurrently disregarded, if we assume the functions $ f_1 $ and $ g_1 $ are approximately equal , the double differential width is 
\begin{eqnarray}
	&\frac{d^2 \Gamma}{d \hat{q^2} dy}*\frac{1}{\Gamma_0}=&12 f_1(\hat{q^2} )\left(y-\hat{q^2}\right) \left(\hat{q^2}-\rho -y+1\right)\;\nonumber\\
	&\frac{d^2 \Gamma}{d \hat{q^2} dx}*\frac{1}{\Gamma_0}=&12 f_1(\hat{q^2} )[ x*(1-x-\rho)]	\;.	
\end{eqnarray}Within a range of $ \hat{q^2} $ that is as narrow as possible, the distribution of individual lepton energies is the same as at the free quark level, differing only by a shape function. If the decay products are positrons and neutrinos, then the double differential widths related to positrons is $ \frac{d^2 \Gamma}{d \hat{q^2} dx} * \frac{1}{\Gamma_0} $.

Taking $\bar{B}^0\to D^{+} +e^{-}+v_{e}$ as an example, we demonstrate the distribution of the widths of double differentials under different $ \hat{q^2} $ values. For ease of comparison, here we uniformly set the coefficient ($6|f_{+}|^{2}$,$12 f(\hat{q^2} )$)to 1.  The maximum $ \hat{q}^2  $ value is $ \hat{q^2}|_{\text{max}}=(1-\sqrt{\rho})^2 $.
%我们以$\bar{B}^0\to  D^{+} +e^{-}+v_{e}$为例，展示了不同取值下双微分宽度的分布，为了便于比较，这里我们将含有形状因子的系数（）统一设定成1。
\begin{figure}
	\centering
	\includegraphics[width=0.7\linewidth]{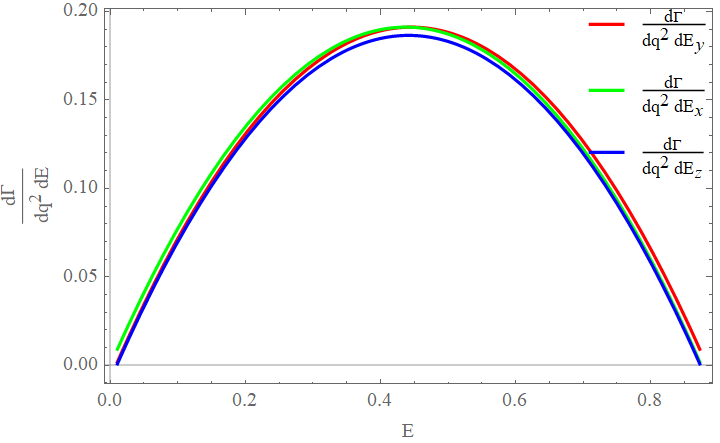}
	\includegraphics[width=0.7\linewidth]{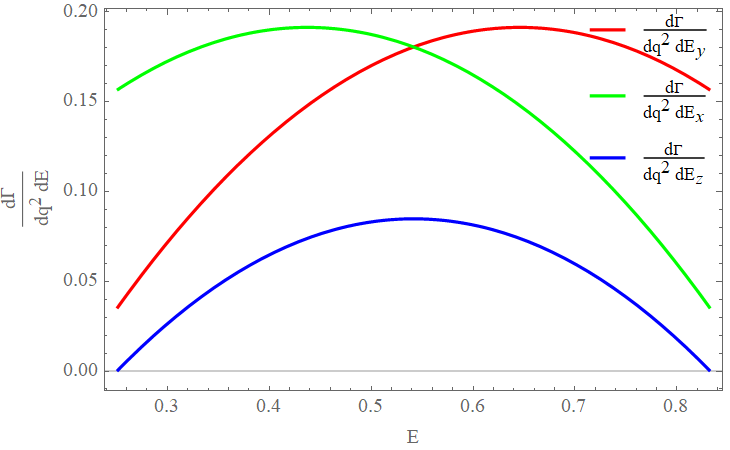}
	\includegraphics[width=0.7\linewidth]{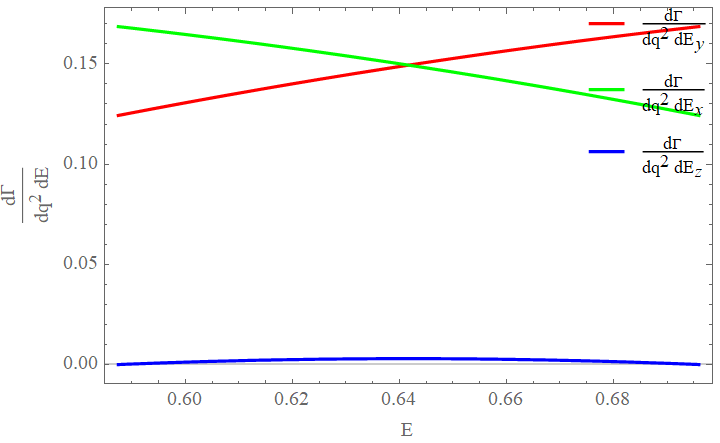}
	\caption{The distribution of electron energy, neutrino energy, and lepton energy at low($\frac{\hat{q^2}|_{\text{max}}}{50}$)\;, middle($\frac{\hat{q^2}|_{\text{max}}}{2}$)\;, high($\frac{49\hat{q^2}|_{\text{max}}}{50}$) $ \hat{q^2} $.}
	%电子能量，中微子能量，轻子能量，在高$ \hat{Q}^2  $处的分布情况}
\label{3q}
\end{figure}
%在q2趋于0的时候，这三种分布趋于相同，当q2逐渐增大的时候，三种分布之间的差别会越来越大。实验上可以通过测量较高处的q2范围内的能量分布而判断哪种理论是最佳的。我们也试了一下当f和g不等的情况，分布大致趋势是一致的。
When $ \hat{q^2} $ approaches 0 in FIG.~\ref{3q}, these three distributions tend to become the same. As $ \hat{q^2} $ gradually increases, the differences between the three distributions will become larger and larger. Experimentally, one can determine which theory is the best by measuring the energy distribution within a higher range of $ \hat{q^2} $. 

We also tried the case where functions $ f_1 $ and $ g_1 $ are not equal,  the double differential width is
\begin{eqnarray}
	&\frac{d^2 \Gamma}{d \hat{q^2} dy}*\frac{1}{\Gamma_0}=12[\frac{(f_1+g_1)^2}{4}\left(y-\hat{q^2}\right) \left(\hat{q^2}-\rho -y+1\right) \nonumber\\&+\frac{(f_1-g_1)^2}{4}y*(1-y-\rho)+\frac{(f_1^2-g_1^2)}{2}\hat{q^2}\sqrt{\rho}]\;\nonumber\\
	&\frac{d^2 \Gamma}{d \hat{q^2} dx}*\frac{1}{\Gamma_0}=12[ \frac{(f_1-g_1)^2}{4}\left(x-\hat{q^2}\right)\left(\hat{q^2}-\rho -x+1\right)\nonumber \\&+\frac{(f_1+g_1)^2}{4}x*(1-x-\rho)+\frac{(f_1^2-g_1^2)}{2}\hat{q^2}\sqrt{\rho}	]\;.	
\end{eqnarray}
and the general trend of the distribution is consistent.	
	\section{Conclusion}
	
This paper presents a unified theoretical framework for calculating exclusive semi-leptonic decays of hadrons, addressing inconsistencies in traditional approaches—particularly in the meson sector.  We introduce a novel method based on the hadronic tensor and Taylor-expanded form factors, which reduces the number of parameters needed to fit experimental data compared to standard Light Cone Sum Rule (LCSR) form factors.  The approach is validated in the baryon sector, where it aligns with existing theory, and applied to meson decays, where it offers a revised parameterization.  The model predicts measurable differences in double differential decay widths, providing a testable way to distinguish it from conventional theories.  The method is also extended to cases where the V-A structure may be violated.  Overall, the framework offers a simpler, more experimentally grounded alternative for analyzing heavy flavor decays.
	%本文提出了计算强子的遍举半轻衰变的统一理论框架，解决了传统方法的不一致性，特别是在介子部分。本文介绍了一种基于强子张量和taylor展开形式因子的新方法，与标准光锥求和规则（LCSR）形式因子相比，该方法减少了拟合实验数据所需的参数数量。该方法在重子扇区得到验证，与现有理论相一致，并应用于介子衰变，在那里它提供了一个修改的参数化。该模型预测了双微分衰变宽度的可测量差异，提供了一种可测试的方法来区分它与传统理论。该方法也推广到V-A结构可能被破坏的情况。总的来说，这个框架提供了一个更简单、更有实验基础的方法来分析重味粒子衰变。

	\section{acknowledgments}
\begin{acknowledgments}
Thanks to the people who gave me different opinions, which made my idea more perfect. This article has been published on the preprint website, arxiv 2401.09065. This research is supported by the institutional research start-up funds(HQ24053).
%	This work is supported by the Natural Science Foundation of China under grant No. XXX. Thanks for some useful discussions with XXX.
\end{acknowledgments}

%	\end{CJK}	
	
\end{document}